\documentclass{article}

\usepackage{PRIMEarxiv}

\usepackage[utf8]{inputenc} 
\usepackage[T1]{fontenc}    
\usepackage{hyperref}       
\usepackage{url}            
\usepackage{booktabs}       
\usepackage{amsfonts}       
\usepackage{nicefrac}       
\usepackage{microtype}      
\usepackage{lipsum}
\usepackage{fancyhdr}       
\usepackage{graphicx}       
\graphicspath{{media/}}     
\usepackage{amsmath,amssymb,amsfonts,amsthm}
\usepackage{algorithmic}
\usepackage{textcomp}
\usepackage{xcolor}
\usepackage{blindtext}
\usepackage{subfig}
\usepackage{cuted}
\usepackage{physics}
\usepackage{xcolor}
\usepackage[long]{optidef}
\usepackage[detect-none]{siunitx}
\usepackage{multirow}
\usepackage{array}
\newtheorem{thm}{Theorem}
\usepackage{makecell}
\title{Transmit Power Optimization of IoT Devices over Incomplete Channel Information\\}
\title{Transmit Power Optimization of IoT Devices over Incomplete Channel Information
\thanks{\textit{This material is based upon work supported by Science Foundation Ireland (SFI) and is co-funded under the European Regional Development Fund under Grant Numbers 13/RC/2077 and 13/RC/2077-P2. \\
\copyright 2023 IEEE. Personal use of this material is permitted. Permission from IEEE must be obtained for all other uses, in any current or future media, including reprinting/republishing this material for advertising or promotional purposes, creating new collective works, for resale or redistribution to servers or lists, or reuse of any copyrighted component of this work in other works.}} 
}

\author{
  Nirmal D. Wickramasinghe \\
  Department of Electronic Engineering \\
  Maynooth University \\
  Ireland\\
  \texttt{nirmal.wickramasinghe.2023@mumail.ie} \\
   \And
  Indrakshi Dey \\
  Walton Institute of Information and Communication Science \\
  South East Technological University \\
  Ireland\\
  \texttt{indrakshi.dey@waltoninstitute.ie} \\
}

\begin{document}
\maketitle

\begin{abstract}
Efficient resource allocation (RA) strategies within massive and dense Internet of Things (IoT) networks is one of the major challenges in the deployment of IoT-network based smart ecosystems involving heterogeneous power-constrained IoT devices operating in varied radio and environmental conditions. In this paper, we focus on the transmit power minimization problem for IoT devices while maintaining a threshold channel throughput. The established optimization literature is not robust against the fast-fading channel and the interaction among different transmit signals in each instance. Besides, realistically, each IoT node possesses incomplete channel state information (CSI) on its neighbors, such as the channel gain being private information for the node itself. In this work, we resort to Bayesian game theoretic strategies for solving the transmit power optimization problem exploiting incomplete CSIs within massive IoT networks. We provide a steady discussion on the rationale for selecting the game theory, particularly the Bayesian scheme, with a graphical visualization of our formulated problem. We take advantage of the property of the existence and uniqueness of the Bayesian Nash equilibrium (BNE), which exhibits reduced computational complexity while optimizing transmit power and maintaining target throughput within networks comprised of heterogeneous devices.
\end{abstract}


\section{Introduction}\label{Introduction}
Internet of Things (IoT) has emerged as a game-changing technology that enables the interconnection of heterogeneous devices and systems. IoT networks present unparalleled prospects for collecting, analyzing, and employing data in real-time, facilitating the ability of making informed decisions grounded on precise information \cite{IoT_overview}. However, resource allocation within IoT networks presents a number of challenges, including issues pertaining to storage, capabilities, spectrum, and energy efficiency (EE), \cite{RA_IoT}. Huawei's white paper \cite{huawei2018global} forecasts a considerable surge in demand for IoT-based services, emphasizing the need for innovative solutions to overcome the challenges associated. IoT devices are typically equipped with constrained energy sources, such as batteries, that have finite capacities. Consequently, the importance of green communication in the advancement of energy-efficient communication and system models for IoT networks has become a critical consideration to overcome energy constraints and sustainability issues \cite{energy_efficient_green_IoT}. Additionally, to combat the excessive energy consumption of IoT devices, a wide variety of architectures featuring policies for energy-efficient data transmission from sensors have been proposed. This diverse range of architectures aims to optimize the energy utilization of IoT networks and ensure their sustainability \cite{Green_IoT_Energy_Saving_Practices_2020}. 

To decrease the power consumption of IoT networks, a plethora of algorithms have been proposed, leveraging various numerical and machine learning (ML) techniques. In their work \cite{IGWO}, the authors have proposed a model to tackle the challenge of energy utilization by addressing non-convex optimization problems in the presence of complex channel states. By utilizing transmit antenna selection and the amplify and forward relaying approach, the authors aim to optimize the system's performance in terms of energy utilization and improve the overall efficiency of the system. And, in a related study by \cite{joint_opt}, a joint optimization strategy has been developed to address the optimization of various parameters in fog computing industrial Internet of Things (IoT) systems. Specifically, the study focuses on optimizing transmission power, offloading ratio, local CPU computation speed, and transmission time. By considering these factors collectively, the aim is to enhance the overall performance and efficiency of fog computing systems in the context of industrial IoT applications.

Furthermore, the allocation of resources for edge computing and heterogeneous IoT applications has been addressed in the literature through the application of reinforcement deep learning techniques. In these studies \cite{RA_Reienforcement_L_01}, the authors aimed to minimize the long-term weighted sum cost, which includes both power consumption and task execution latency. Additionally, \cite{RA_Reienforcement_L_02} aimed to maximize the quality of experience (QoE) for the users. To achieve these objectives, the authors utilized Q-value approximation methods within the reinforcement learning framework. By employing this approach, they sought to optimize resource allocation decisions, considering the trade-off between power consumption, task execution latency, and user QoE. However, traditional machine learning techniques, often employed for optimization and pattern recognition tasks, primarily rely on data-driven approaches. These methods typically require access to comprehensive information about the entire network, which may prove challenging and resource-intensive, particularly in large-scale networks with a significant number of nodes. As the network size grows, the complexity of these machine-learning techniques escalates exponentially, leading to resource inefficiencies. The computational and memory requirements can become prohibitive, making the application of such techniques impractical or even infeasible in certain scenarios. As a result, researchers are actively exploring alternative approaches to tackle these challenges and enhance the efficiency of optimization and pattern recognition tasks in large-scale networks.

Compared to numerical and learning-based techniques, game-theoretic solutions offer several advantages in addressing optimization problems \cite{Game_Theory_IoT_survey}. One key advantage is their ability to capture and utilize a comprehensive understanding of the strategic dynamics among entities involved in the optimization process. By considering the strategic behavior of entities, game-theoretic solutions can capture the complex dynamics of the optimization problem more accurately. Therefore, game-theoretic solutions offer a promising avenue for addressing optimization problems in a wide range of domains, including IoT networks. The challenge of selecting joint access and resource allocation in the unmanned aerial vehicle (UAV) assisted IoT networks is addressed in \cite{Stakelberg_GT} using the non-cooperative Stackelberg game theory. Authors in \cite{Bayesian_Stakelberg_GT} explore a combination of Stackelberg and Bayesian methodologies to enhance utility optimization and corresponding strategy space in heterogeneous networks. Furthermore, \cite{Bayesian_radar} suggests a non-cooperative game theoretic approach employing Bayesian strategies to address the issue of resource allocation in multi-static radar networks. In addition, \cite{Bayesian_batterylessSystem} employs game theory to investigate the operations of energy harvesting for battery-less powering of IoT devices drawn from multiple sources, which are typically constrained in terms of energy provision and rely exclusively on harvesting procedures. Although game theory has been sporadically leveraged in literature, it has never been adequately exploited for energy optimization in IoT new despite its numerous advantages, game theory has been underutilized in the literature when it comes to energy optimization in IoT networks. The potential of game theory in this context has not been fully realized, and there is a need for more comprehensive exploration and exploitation of its capabilities.
 
 This paper introduces an approach that embraces Bayesian game theory to minimize the transmit power among IoT nodes, leveraging incomplete channel information available among nodes in each instance. The proposed approach is resilient to maintain the channel throughput in the presence of interference among individual IoT nodes, thereby ensuring reliable communication. In this work, we provide a detailed exposition of the rationale behind the employment of Bayesian game theory. Subsequently, we present a graphical representation elucidating the existence and uniqueness of the Bayesian Nash Equilibrium (BNE), highlighting its potential to significantly reduce computational complexity and enhance robustness in the context of complex IoT networks. Furthermore, the proposed method's optimal power strategies precisely converge to brute-force results and outperform conventional artificial neural network (ANN) approaches. Finally, a comprehensive discussion is conducted on the ability of Bayesian game theory techniques to analyze transmit signal interactions, facilitated by an intermediate stage of the algorithm.

\section{System Model}\label{sec:System_Model}
\begin{figure}[t]
    \centering
    \includegraphics[width=\textwidth]{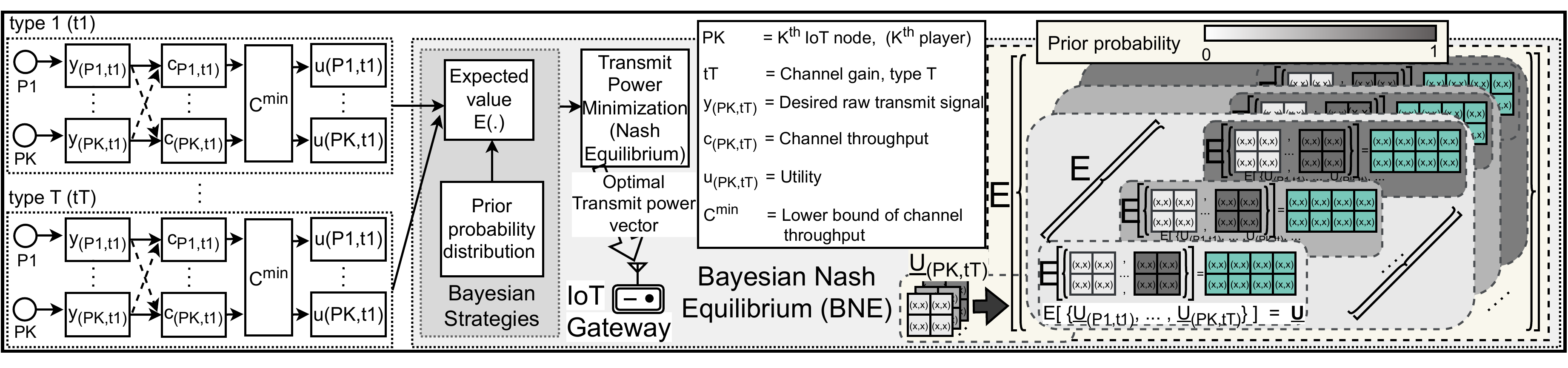} 
    \caption{Representation of Bayesian Game model based simulator for power optimization of nodes in massive IoT network.}
    \label{fig:sys_model}
    \end{figure}

In this paper, a simple IoT  transmission system model is represented from $K$ number of IoT nodes connected with a gateway as shown in \figurename~\ref{fig:sys_model}. IoT nodes with diverse features can be sensed by multiple transceivers within a telecommunication infrastructure. These nodes operate under the assumption of static deployment and self-awareness of their individual locations. Here, IoT nodes refer to transmitters that convey information over a random wireless channel which depends on the existing environmental conditions of the receiver gateway. We can mathematically formulate the received signal from the $k^{th}$ IoT node as 
    \begin{align}\label{rx_sig}
    y(t) = h_{k}(t) x_{k}(t) + \sum_{j=1, j\neq k}^{K} h_{j}(t) x_{j}(t) + n(t) 
    \end{align}
where $x_{k}(t)$, $h_{k}(t) \in \mathbb{C}$ are the transmitted signal and the small-scale channel fading vector respectively. $h_{k}(t) x_{k}(t)$ is the desired signal, $h_{j}(t) x_{j}(t)$ is the interference signal from the rest of IoT nodes and the last term can be identified as the zero-mean additive white Gaussian noise, $n(t) \sim \mathcal{N}(0,\,\sigma_{n}^{2})$. 

In this problem, the coherence time of the channel is smaller than the delay spread of the transmitted signal due to the small-scale and multi-path fading environment. Therefore, the desired signal of $k^{th}$ IoT node may have interfered with the receiving signals of the rest of the IoT nodes. As a result, we can derive the lower bound on the normalized achievable rate which is known as the throughput of the stationary channel as a function of signal-to-interference noise ratio (SINR) as 
    \begin{align}\label{Eq:cap_x}
		c_{k}& = B\log_2\bigg(1+ \frac{|{h}_k{x}_k|^2}{\sum_{j=1, j\ne k}^{K}|{h}_j{x}_j|^2+\sigma_{n,k}^2}\bigg)
	\end{align}
 where $SINR_{k}=\frac{|{h}_k{x}_k|^2}{\sum_{j=1, j\ne k}^{K}|{h}_j{x}_j|^2+\sigma_{n,k}^2}$ and $\sigma_{n,k}^2$ is the noise power of IoT node $k$.
In addition, the transmitted signal $x_{k}(t)$ can be re-expressed as
\begin{align}
    x_{k}(t) = \sqrt{p_{k}(t)} \cdot s_{k}(t)
\end{align}
where $p_{k}(t)$ and $s_{k}(t)$ are the transmit power and zero-mean data symbol of $k^{th}$ node at time $t$. The modulation of data symbols onto the $I_{DC}^k$ picked up from the DC transmission current vector, $\boldsymbol{I}_{DC, K} = [I_{DC}^1,\ldots, I_{DC}^{K}]$, which is proportional to the power vector of $\sqrt{\boldsymbol{P_{K}}}$ establishes reliable communication. Suppose $s_k\in\{\pm1\}$ as the binary data symbol intended for the $k^{th}$ IoT node, and $\boldsymbol{s}=[s_1,\ldots,s_{K}]$ is the symbol vector. For the sake of independent and identical distribution, $i.i.d.$ of the transmission, the covariance matrix of $\boldsymbol{s}$ is $\mathbf{C}_s=\mathbf{I_{K}}$ where $\mathbf{I_{K}}$ is the identity matrix with the size of $K$. Thus, the throughput can be further re-formulated in terms of power $p_{k}$, as
    \begin{align}\label{Eq:cap_p}
		c_{k}& = B\log_2\bigg(1+ \frac{|{h}_k|^2{p}_k}{\sum_{j=1, j\ne k}^{K}|{h}_j|^2{p}_j+\sigma_{n,k}^2}\bigg)
	\end{align}
 The channel throughput of node $k$ is dependent on the CSI of each neighbor included in the IoT architecture. Therefore, the joint probability distribution of individual channel distributions of each IoT node is responsible for the variation in channel throughput, which can be used as a performance metric to analyze the interaction among various transmit signals of IoT nodes.
 
\section{Transmit Power Optimization}\label{sec:Optimization_Prob}
In this section, we formulate the optimization problem aiming to minimize the  objective function of the total transmit power
	\begin{mini!}
    	{\boldsymbol{p}_{k}}{\sum_{k=1}^{K} p_{k}(h_{k})}
    	{\label{opt_pb}}{}
    	\addConstraint{}{C1: c_k\ge C^{\min}, \quad \forall k\in \mathcal{K}} \label{C1}
    	\addConstraint{}{C2: p_{k}(h_{k})\ge p_{k}^{\min}, \quad \forall k\in\mathcal{K}} \label{C2}
    	\addConstraint{}{C3: p_{k}(h_{k})\le p_{k}^{\max}, \quad \forall k\in\mathcal{K}}, \label{C3}
        \addConstraint{}{C4: \eqref{Eq:cap_p}} \label{C4}
    \end{mini!}
subject to the limitations of \ref{C1},\ref{C4} which stipulates a minimum channel capacity, and \ref{C2},\ref{C3} are the lower and upper bounds of transmit power requirements. Furthermore, $c_{k}$ is the channel throughput experienced by $k^{th}$ IoT node, and the power strategy of $p_{k}(h_{k})$ is a function of channel gain from individual $K$ number of IoT devices. Then, the boundary values for throughput and power can be defined as $C^{\min}$ and $p_{k}^{\min}$,  $p_{k}^{\max}$ respectively.

The derived optimization problem is validated under the case of IoT node performance in the transmission system with knowing all required information. To be more precise, at time t, the fellow channel gain of ${h_{1}(t), \ldots, h_{K}(t)}$ are common knowledge for all $K$ IoT transmitters. The convex nature of the hypothetical single-node power minimization problem with complete information ~\cite{boyd2004convex}, results in the optimal power being derived from a linear programming problem. 
For convenience, Lagrangian duality principles can be used to solve the problem \eqref{opt_pb} and therefore, the augmented objective function is

    \begin{align}
    L(p_{k}, \lambda_{i}, s_{i}) = p_{k}-\lambda_{1}(c_{k}-C^{\min}-s_{1}^{2})\\
      -\lambda_{2}(p_{k}-p_{k}^{\min}-s_{2}^{2})\nonumber\\
      -\lambda_{3}(p_{k}-p_{k}^{\max}+s_{3}^{2})\nonumber 
    \end{align}
    
    
    where, $\lambda_{i}$ and $s_{i}$ are the Lagrangian multipliers and slack variables respectively, and $i=1,2,3$. Then, the gradient operates on the three variables as follows
    \begin{align}\label{Eq:Lag_grad}
        \pdv{L}{p_{k}}& = 1-\lambda_{1}\pdv{c_{k}}{p_{k}}-\lambda_{2}-\lambda_{3}\\
        \pdv{L}{\lambda_{1}}&=-(c_{k}-C^{\min}-s_{1}^{2})\\
        \pdv{L}{\lambda_{2}}&=-(p_{k}-p_{k}^{\min}-s_{2}^{2})\\
        \pdv{L}{\lambda_{3}}&=-(p_{k}-p_{k}^{\max}+s_{3}^{2})\\
        \pdv{L}{s_{i}}&=\pm2 \lambda_{i} s_{i}
    \end{align}
After avoiding all infeasible scenarios from the bunch of possible complementary slackness conditions, the optimal power of node $k$ can be derived as follows
\begin{align}\label{Eq:Lag_obj_sol}
\pdv{L}{p_{k}}&= 0\Leftrightarrow\lambda_{k}= \frac{1}{\pdv{c_k}{p_{k}}}\\
\lambda_{k}& = \frac{1}{B|{h}_k|^2}\left(\sum_{j=1, j\ne k}^{K}|{h}_j|^2{p}_j+|{h}_k|^2{p}_k+\sigma_{n,k}^2\right)\\
\lambda_{k}& = \frac{1}{B|{h}_k|^2}\left(\sum_{j=1}^{K}|{h}_j|^2{p}_j+\sigma_{n,k}^2\right)
\end{align}
where the duality multiplier $\lambda_{k}$ is opted such that the satisfaction of equality of the power constraints in \eqref{opt_pb}. However, the derived optimum power of $k^{th}$ node depends on the power strategies of the neighbors which node $k$ doesn't know, and vice versa. Therefore, it is required to analyze instantaneous power decisions under interactions among IoT nodes and their respective types.


\section{Game theory based solution}\label{Game_model}
Game theory is a remarkable mathematical framework to analyze the strategic interactions among users in a decision-making process. It provides a way to model and predict how rational individuals will behave when their actions affect the outcomes for themselves and others.

\subsection{Game model}\label{Game_model}
\begin{figure}[h]
        \centering
        \includegraphics[width=0.5\textwidth]{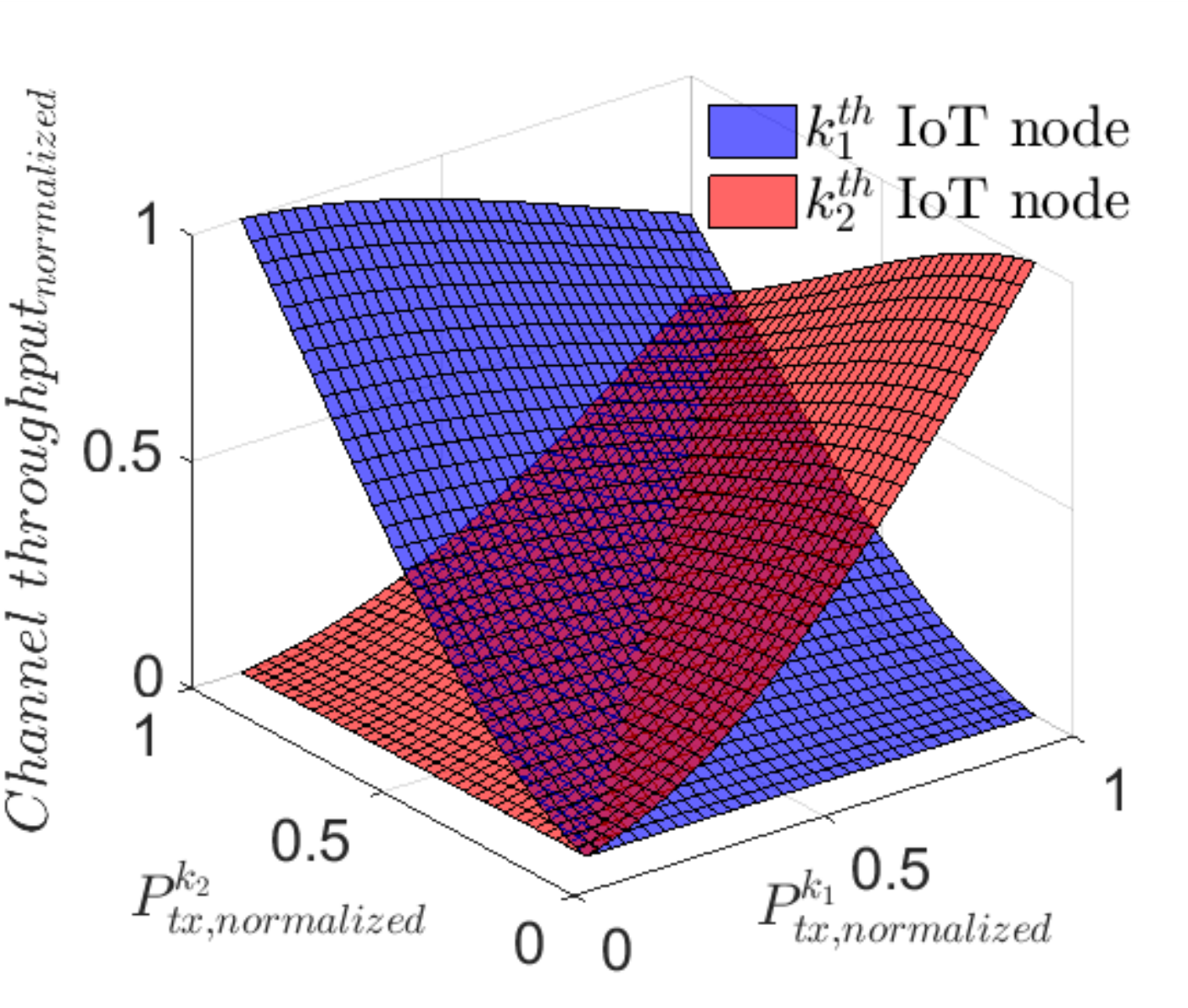}
        \caption{Variation $(Channel$ $throughput_{normalized} \text{ v/s } P_{tx, normalized}^{k})$ of the 
        normalized channel throughput with Transmit Power strategies of IoT nodes $k_{1}$ and $k_{2}$.}
        \label{fig:throughput}
    \end{figure}

The stated problem up to now can be defined using the standard representation of a game model which is known as a normal form game, or a game in the strategic form: 
    \begin{align}\label{game_model_1}
        \mathcal{G} \triangleq \langle \mathcal{K}, \mathcal{A},  \mathcal{U} \rangle 
    \end{align}
    where 
    \begin{itemize}
        \item The set of IoT nodes is $\mathcal{K}=\{1,\ldots,k, \ldots, K\}$
        \item Node $k$ has a set of actions, $a_{k}$ which are generally referred to as pure strategies. The cross product of action space $\mathcal{A}=\times_{k \in \mathcal{K}}a_{k}$ be the set of all profiles of actions with a generic element denoted by $\textbf{a}=(a_{1}, \ldots, a_{K})$.
        \item User $k$ 's payoff or utility as a function of $\mathcal{A}$ taken is described by a function of: $\mathcal{U}:\mathcal{A} \rightarrow \mathbb{R}$ where $u_{k}(\textbf{a})$ is the IoT node $k$ 's payoff if $\textbf{a}$ is the profile of actions in the IoT pool.
        \item The utility set: $\mathcal{U}=\{u_{1},\ldots, u_{K}\}$ such that $u_{k}$ is derived as follows.
    \end{itemize}
Let's rewrite the channel throughput for user $k$ with respect to the user interaction among each other which means the range of strategic interactions between user $k$ and others.

    \begin{subequations}\label{eq:cap_user_interaction}
    \begin{equation}
		C_{k} = B\log_2\left(1+ \frac{(|h_{k}|^2p_{k})}{\sum_{\forall{-k \in K}}|h_{-k}|^2p_{-k}+\sigma_{n,k}^2}\right).
    \tag{\ref{eq:cap_user_interaction}}
    \end{equation}
    where 
    \begin{align}\label{eq:rest_users}
    h_{-k}(t) = \{h_{1}(t),\ldots, h_{k-1}(t), h_{k+1}(t),\ldots, h_{K}(t)\}\\
    p_{-k}(t) = \{p_{1}(t),\ldots, p_{k-1}(t), p_{k+1}(t),\ldots, p_{K}(t)\}
    \end{align}
    are the set of channel realization and corresponding power set for all IoT nodes plus neglecting $k^{th}$. 
    \end{subequations}
When considering the variation of channel throughput versus transmit power for two IoT nodes, as shown in \figurename~\ref{fig:throughput}, it can be observed that the channel throughput increases with higher transmit power from each IoT node. Despite the fact that the $k_{2}^{th}$ IoT node acts as an interfering signal for the desired signal of the $k_{1}^{th}$ IoT node, the overall channel throughput effectively increases as both transmit power vectors in each node are incrementally raised and vice versa. In accordance with the objective function of the proposed optimization problem stated in \eqref{opt_pb}, the utility function aims to select and reflect the level of satisfaction of the IoT node based on the minimum desired throughput. This utility function takes into account the IoT node's preferences and satisfaction level, and it should serve as a criterion for selecting the minimum transmit power vector that maximizes the desired minimum throughput. Therefore, a suitable mapping function from the channel throughput to the utility function can be selected as follows.

 .
    \begin{subequations}\label{eq:utility_fn}
    \begin{equation}
        u_{k}(t)=
    \begin{cases}
        \frac{1}{\Delta c_{k}(t)+\epsilon}, & \text{if $c_{k}(t)\geq 0$}.\\
        \Delta c_{k}(t), & \text{otherwise}.
    \end{cases}
    (\epsilon \ll) \in \mathbb{R^{+}}
    \tag{\ref{eq:utility_fn}}
    \end{equation}
    where any arbitrary throughput of the channel reaching towards the threshold conditions is categorized as
    \begin{align}\label{eq:th_satisfaction}
        \Delta c_{k}(t) &= c_{k}(t) - C^{\min} \in \mathbb{R}
    \end{align}
    \end{subequations}
However, the term "incomplete information of channel" pertains to a scenario wherein each IoT node possesses insufficient knowledge or comprehension pertaining to the characteristics, preferences, or intentions of the other nodes. This can lead to uncertainty and sub-optimal outcomes. 

\subsection{Game model for Incomplete Information}\label{incomplete_info}

In this paper, information incompleteness can be defined formally such that, the node $k$ does not have knowledge of channel state information of the rest as explained in \eqref{eq:rest_users}. The analysis of incomplete information games is more complex than that of complete information games because it requires taking into account users' beliefs and information, which may be updated as the game unfolds. Although such uncertainty is applicable to all users in the network, every user knows his own information or types itself, and the probability distribution of a set of types is common knowledge among all users. The introduced different types and respective  mutual beliefs are important features in a game with incomplete information to re-establish the game model which is equivalent to Harsanyi's game model \cite{Harsanyis_GT}, \cite{maschler_solan_zamir_2013} as

    \begin{align}\label{game_model_1}
        \mathcal{G} \triangleq \langle \mathcal{K}, (\mathcal{T}_{k})_{k \in \mathcal{K}}, \mathcal{P}, \mathcal{S}, (s_{t})_{t \in \times_{k \in \mathcal{K}}\mathcal{T}_{k}}  , \mathcal{U} \rangle 
    \end{align}
    where
    \begin{itemize}
        \item $\mathcal{K}$ is a finite set of IoT devices and $k \in \mathcal{K}$.
        \item $\mathcal{T}_{k}$ is a finite set of types for node k, for each $k \in \mathcal{K}$, and the set of type vectors are denoted by $\mathcal{T}= \times_{k \in \mathcal{K}}\mathcal{T}_{k}$.
        \item $\mathcal{P} \in \triangle(\mathcal{T})$ is a probability distribution over the set of type vectors, $\mathcal{T}_{-k}= \times_{j \neq k} \mathcal{T}_{j}$ and $t_{-k}= (t_{j})_{j \neq k}$ that satisfies $p({t_{k})} \coloneqq \sum_{t_{-k} \in \mathcal{T}_{-k}} p(t_{k}, t_{-k}) > 0, \forall k \in \mathcal{K}, \forall t_{k} \in \mathcal{T}_{k}$.
        The probability distribution of prior belief is characterized using the fast-fading channel conditions as
        \begin{align}\label{Rayleigh_dis}
            p(t_{k})=\frac{t_{k}}{R_{coff}^2}\exp(- \frac{t_{k}^{2}}{2R_{coff}^2})
        \end{align}
        which is known as the probability density function of Rayleigh distribution.
        \item $\mathcal{S}$ is a set of states or strategies of IoT system, and every state of the system $s \in \mathcal{S}$ is a vector $s=(\mathcal{K}, (A_{k})_{k \in \mathcal{K}}, (u_{k})_{k \in \mathcal{K}})$, where $A_{k} \in \mathcal{A}$ is a nonempty set of actions of node $k$ and $u_{k}: \times_{k \in \mathcal{K}}A_{k} \rightarrow \mathbb{R}$ is the payoff function of node k$.$
        \item $s_{t}=(\mathcal{K}, (A_{k}(t_{k}))_{k \in \mathcal{K}}, (u_{k}(t))_{k \in \mathcal{K}}) \in \mathcal{S}$ is the strategy profile for type vector $t$, $\forall t \in \mathcal{T}$. Thus, node $k$'s action set in the state $s_{t}$ depends on the $t_{k}$ itself only and is independent of the types of the other nodes.
    \end{itemize}

        \begin{table}[!h]
        \begin{center}
        \begin{tabular}{|c|c|c|}
        \hline
        \shortstack{\newline \\ \newline \\ \textbf{$P_{k}|P_{k+1}$} \\ \newline} & \shortstack{\textbf{$A_j$} \\ \newline} & \shortstack{\textbf{$A_{j+\Delta j}$} \\ \newline}\\
        \hline
        \shortstack{\textbf{$A_{i}$} \\ \newline \\ \newline \\ \newline}  & \shortstack{\newline \\ \newline \\ \textbf{\textit{$\left[(u_{P_{k}}^{t})_{A_{i}, A_{j}},(u_{P_{k+1}}^{t})_{A_{i}, A_j}\right]$}} \\ \newline \\ \textbf{\textit{$ \rightarrow r_{A_{i}, A_j}$}} \\ \newline} & \shortstack{\newline \\ \newline \\  \textbf{\textit{$\left[(u_{P_{k}}^{t})_{A_{i}, A_{j+\Delta j}},(u_{P_{k+1}}^{t})_{A_{i}, A_{j+\Delta j}}\right]$}} \\ \newline \\ \textbf{\textit{$ \rightarrow r_{A_{i}, A_{j+\Delta j}}$}} \\ \newline} \\
        \hline
        \shortstack{\textbf{$A_{i+\Delta i}$}\\ \newline \\ \newline \\ \newline}  & \shortstack{\newline \\ \newline \\ \textbf{\textit{$\left[(u_{P_{k}}^{t})_{A_{i+\Delta i}, A_j},(u_{P_{k+1}}^{t})_{A_{i+\Delta i}, A_j}\right]$}} \\ \newline \\ \textbf{\textit{$ \rightarrow r_{A_{i+\Delta i}, A_j}$}} \\ \newline}  & \shortstack{\newline \\ \newline \\ \textbf{\textit{$\left[(u_{P_{k}}^{t})_{A_{i+\Delta i}, A_{j+\Delta j}},(u_{P_{k+1}}^{t})_{A_{i+\Delta i}, A_{j+\Delta j}}\right]$}} \\ \newline \\ \textbf{\textit{$ \rightarrow r_{A_{i+\Delta i}, A_{j+\Delta j}}$}} \\ \newline} \\
        \hline
        \end{tabular}
        \vspace{5mm}
        \caption{Game matrix for type t / $G_{t}$.}
        \label{G_mat}
        \end{center}
        \end{table}

         \begin{multline}\label{utility}
         \{(u_{P_{k}}^{t})_{A_i, A_{j+\Delta j}} \lesseqgtr (u_{P_{k+1}}^{t})_{A_{i+\Delta i}, A_j} < (u_{P_{k}}^{t})_{A_i, A_j} \lesseqgtr (u_{P_{k+1}}^{t})_{A_i, A_j} < (u_{P_{k}}^{t})_{A_{i+\Delta i}, A_{j+\Delta j}} \}\\
         \cup \{\lesseqgtr (u_{P_{k+1}}^{t})_{A_{i+\Delta i}, A_{j+\Delta j}} < u_{P_{k}}^{t})_{A_{i+\Delta i}, A_j} \lesseqgtr (u_{P_{k+1}}^{t})_{A_i, A_{j+\Delta j}}\}
        \end{multline}
        
        \begin{align}\label{reward}
            r_{A_i, A_{j+\Delta j}} \lesseqgtr r_{A_{i+\Delta i}, A_j} < r_{A_i, A_j} < r_{A_{i+\Delta i}, A_{j+\Delta j}}            
        \end{align}

    \begin{figure}[t]
        \centering
        \includegraphics[width=0.7\textwidth, trim={2.6cm 4.1cm 1.4cm 2.9cm},clip]
        {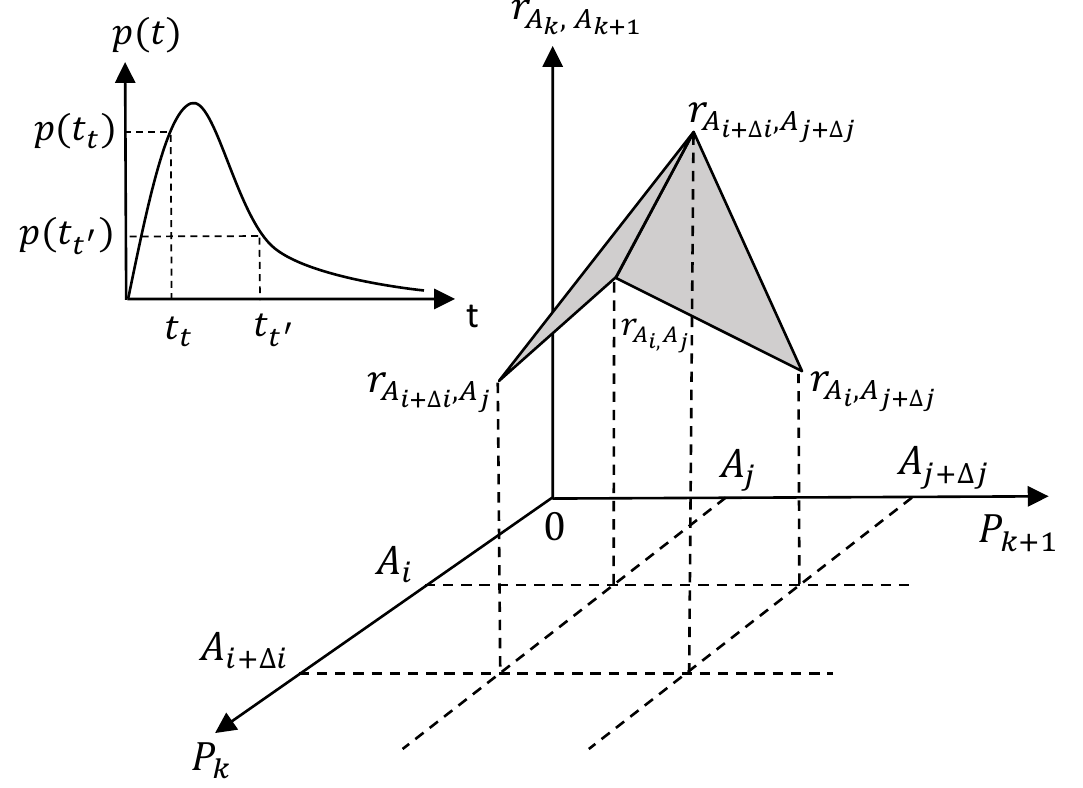} 
        \caption{Overview of interconnection between Game model and Objective function (utility based generalized reward function) of desired optimization problem with respect to two consecutive transmit power actions of IoT nodes $k$ and $k+1$, based on a random prior belief of $p(t_{t})$ at type $t$.}
       \label{fig:overview}
    \end{figure}
    
By assuming the channel distribution is vary under the Rayleigh fading distribution in \eqref{Rayleigh_dis}, let's pick up a random type of channel state $t_{1}$ with the corresponding probability of $p(t_{1})$. Then, the payoff matrix for IoT node $k$ and $k_{2}$, regarding the two consecutive actions is presented in table \ref{G_mat}. Here, a reward can be offered for IoT devices according to the equilibrium state of each power strategy themselves. Assuming that the sample game model has been calibrated using the Prisoner's Dilemma game model \cite{prisonnersDil}, as inferred from analogies presented in equations \ref{utility} and \ref{reward}, the graphical representation of the feasible region is depicted in figure \ref{fig:overview}. The employment of a game matrix is a valuable method for examining strategic interactions between nodes, and it can assist in identifying the most advantageous power strategy for each node based on their adversary's decisions. 
This is achieved through the implementation of diverse solution concepts, including the Nash equilibrium \cite{NashEq}, which determines the strategy combination in which neither player has the motive to alter their strategy. In this paper, it is crucial to contemplate the game model for a range of IoT devices with several types encompassing different channel states, within a unified illustration, instead of limiting the analysis to a single arbitrary type.

\subsection{Bayesian Strategies}\label{bayesian_stretegies}

Games that entail incomplete information can be examined through the utilization of a conditional probability method called the Bayesian strategy. The equilibrium state of Bayesian strategies can be depicted utilizing the following theorem which was proofed in \cite{Harsanyis_GT}.

\begin{thm}[Harsanyi, 1967]\label{Harsanyis_thm}
In a game with incomplete information in which the number of types of each player or IoT node is finite, every Bayesian equilibrium is also a Nash equilibrium, and conversely vice versa.
\end{thm}

In other words, no IoT device has a profitable power deviation after it knows which type it is if and only if the device has no profitable power deviation before knowing the type itself. Recall that in the definition of the game model  with incomplete information \eqref{game_model_1}, required that $p(t_{k})>0$ for each node $k$ and each type $t_{k} \in \mathcal{T}_{k}$. To reflect the Bayesian approach, it is essential to define the conditional expected payoff in the game at the interim stage for a random node $k$ as
\begin{align}
    \bar{U}_{k}(s|t_{k}) \coloneqq \sum_{t_{-k}\in \mathcal{T}_{-k}}p(t_{-k}|t_{k})U_{k}((t_k,t_{-k});s)
\end{align}
where
\begin{align}
    p(t_{-k}|t_{k})=\frac{p(t_{k},t_{-k})}{\sum_{t^{'}_{-k}\in \mathcal{T}_{-k}}p(t_{k},t{-k}^{'})}=\frac{p(t_{k},t_{-k})}{p(t_{k})}
\end{align}
Then the optimal power strategy vector $s^{*}=(s_{1}^{*}, s_{2}^{*}, \dots , s_{k}^{*})$ is a Bayesian equilibrium if for each player $k \in \mathcal{K}$, each type $t_{k}\in \mathcal{T}_{k},$ and each possible action $a_{k}\in\mathcal{A}_{k}(t_{k})$,
\begin{align}
    \bar{U}_{k}(s^{*}|t_{k}) \geq \bar{U}_{k}((a_{k}, s_{-k}^{*})|t_{k})
\end{align}
    
Another ambiguous advantage of formatting the power optimization problem using Bayesian strategies is the property of obtaining a unique decision strategy, proofed in \cite{MAC_fading_Bayesian}

\begin{thm}\label{exist_uniq}
There exists a unique Bayesian equilibrium strategy in the proposed finite $\mathcal{K}$ user Bayesian game model.
\end{thm}

Theorem \ref{exist_uniq} is experimentally investigated using the simulation results in \figurename~\ref{fig:exist_unique} with specifying the problem between two Iot nodes. The presence and exclusivity of the Bayesian equilibrium, confirmed through the intersection point of each node's optimal power response, further reinforce the efficiency of the algorithm. The peaks in the illustration represent the advantageous power deviation of individual nodes, while any deviation from the intersection point of best responses should be avoided.

\section{ANN-based approach}\label{benchmarks}
    \begin{figure}[h]
        \centering
        \includegraphics[width=0.7\textwidth, trim={2.8cm 5.1cm 3.5cm 6.6cm},clip]{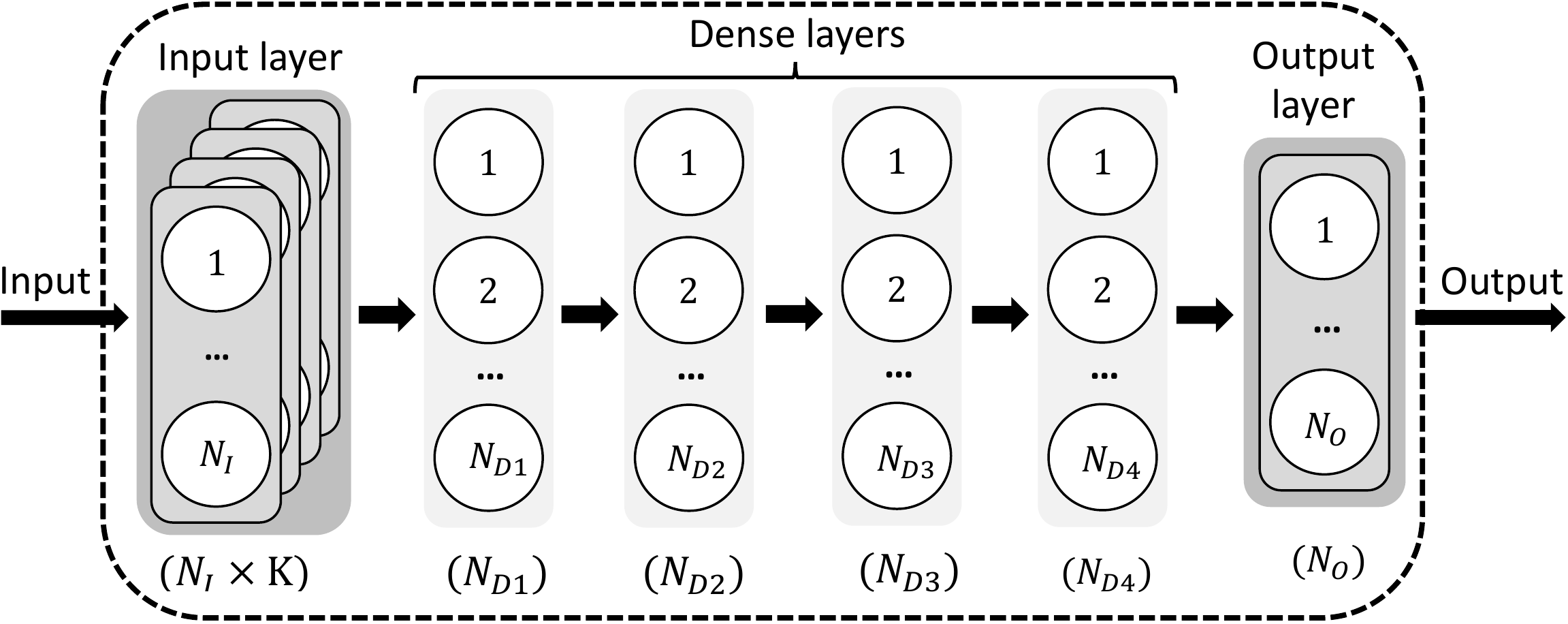}
        \caption{ANN Architecture of parameter optimization}
        \label{fig:ANN_diag}
    \end{figure}
    
To address the optimization problem at hand, our attention is directed towards a lightweight ANN model illustrated in \figurename~\ref{fig:ANN_diag}, which serves as a benchmark for comparison purposes. In this block, the  optimal power strategy is predicted from the inputs of incomplete channel states. To accomplish the objective, we have implemented a regression model that consists of four sequential dense layers, each comprising $N_{D(n))}$ nodes. The outputs of each dense layer are denoted as

\begin{align}
    \hat{\mathbf{x}}^{D(n)}=\sigma_{RELU}\left(\mathbf{W}_{D(n-1)}\hat{\mathbf{x}}^{D(n-1)} + \mathbf{k}_{D(n-1)}\right)
\end{align}
Then the output vector is
\begin{align}
    \hat{\mathbf{y}}=\sigma_{linear}\left(\mathbf{W}_{D(N)}\hat{\mathbf{x}}^{D(N)} + \mathbf{k}_{D(N)}\right)
\end{align}
where $\hat{\mathbf{x}}^{D(n)} \in \mathbb{R}^{N_{D(n)}}$ then, $\mathbf{W}_{D(n)} \in \mathbb{R}^{N_{D(n-1)\times N_{D(n)}}}$, are the weight vectors and $\mathbf{k}_{D(n)} \in \mathbb{R}^{N_{D(n)}}$ are the bias vectors under $n=\{x \mid 1<x\leq N \} \in \mathbb{Z}^{+}$. In order to obtain optimal power values under incomplete channel information, a water-filling-based brute force approach was utilized on the training dataset. The training procedure begins with an initial condition of $\theta$, followed by the application of forward propagation to generate the output vector $\hat{\mathbf{y}}$. For the parameter optimization, the mean square error (MSE) between $\hat{\mathbf{y}}$ and the optimum parameter vector $\mathbf{y_{0}}$, is utilized as the loss function which is expressed as
\begin{align}
    \phi(\mathbf{\theta}) = \frac{1}{M}\sum_{m=1}^{M}\vert \hat{\mathbf{y}}- \mathbf{y_0}\Vert^2 
\end{align}

In the given context, $M$ refers to the mini-batch size employed during the training process. And, the $\theta$ is updated in each batch using the stochastic gradient descent algorithm, which can be expressed as $\mathbf{\theta}^+:=\mathbf{\theta} - \epsilon \nabla\phi(\mathbf{\theta})$, where $\epsilon$ represents the learning rate.

\section{Numerical Results and Discussion}
    \begin{figure}[t]
        \centering
        \includegraphics[width=0.7\textwidth]{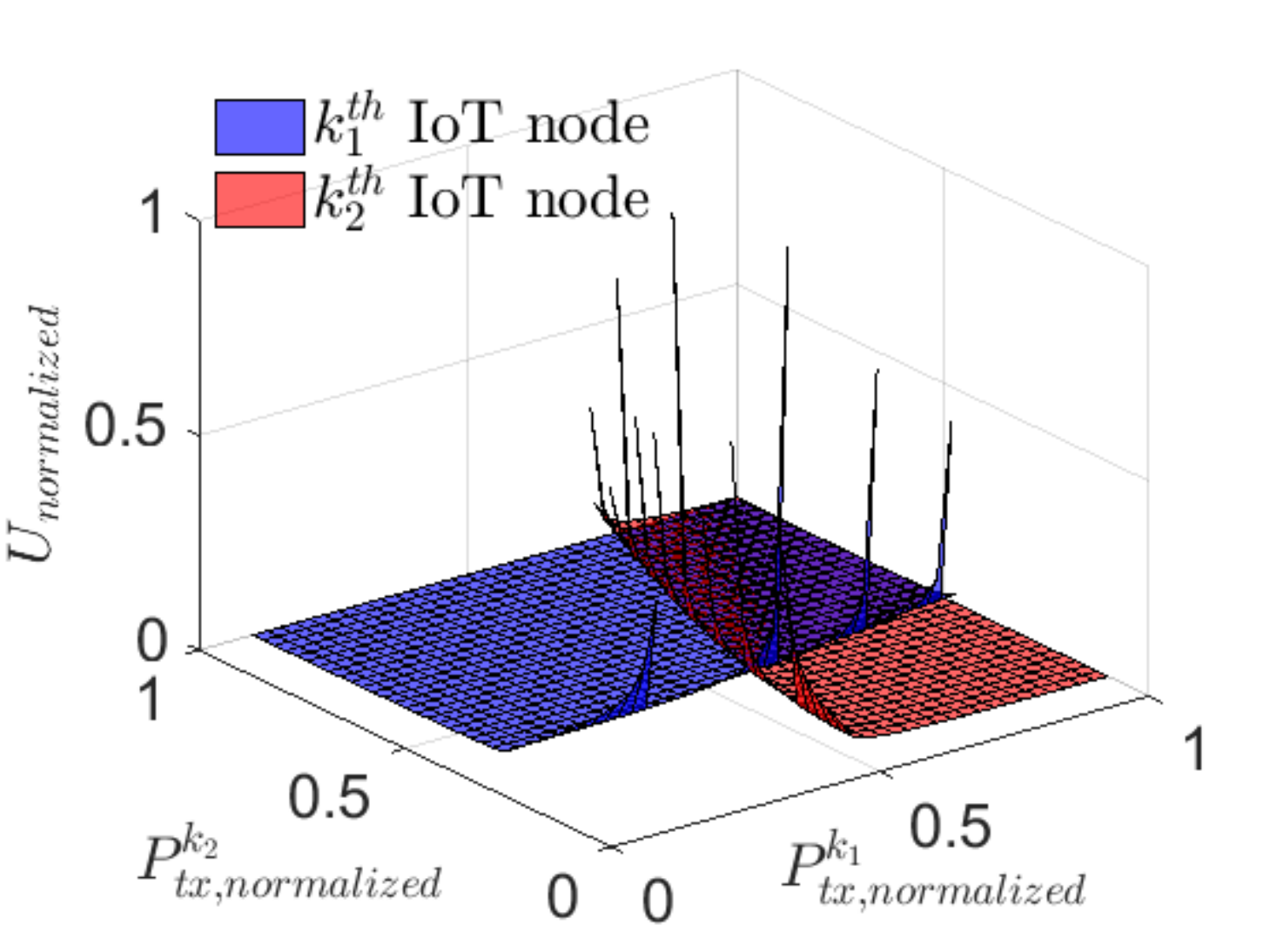}
        \caption{Variation $(U_{normalized} \text{ v/s } P_{tx, normalized}^{k})$ of the 
        normalized Utility function with Transmit Power strategies of IoT nodes $k_{1}$ and $k_{2}$, as graphical illustrative evidence for the property of Existence and Uniqueness of BNE.}
        \label{fig:exist_unique}
    \end{figure}
This section will discuss the impact of the variations of physical parameters on Bayesian strategies using simulation results. We consider normalized channel gains and corresponding transmit power values to analyze performances under different scenarios. In this preliminary study, the Bayesian game is built as a two-player-two-type model $\{k_{1}, k_{2}\} \times \{t_{1}, t_{2}\}$ to simulate the proposed approach and performance. IoT nodes $k_{1}$ and $k_{2}$ experience a probabilistic variation among different channel states or types. The channel gains are Rayleigh distributed with a Rayleigh coefficient of $R_{coff} = 0.5$. The noise power variance is set to $\sigma_{n,k}^2 = 0.1$. 

\figurename~\ref{fig:exist_unique} demonstrates the utility obtained by using channel throughput, as defined in \eqref{eq:utility_fn}, against the transmit (Tx) power strategies of IoT nodes $k_{1}$ and $k_{2}$. The peaks in the graph indicate the optimal transmit power deviation of individual nodes, referred to as the best response of each IoT device. However, the novelty of the proposed approach lies in its capability to bypass individual selfishness and converge towards the equilibrium point of power strategies, as denoted by the intersection point of the best responses. 
Therefore, the proposed algorithm can employ iterative elimination of rows and columns that correspond to the strictly dominated power strategies of the game matrix. As a consequence, a significant reduction in computational complexity can be achieved, going down from the computational order of $\mathcal{O}({\lvert \mathcal{A} \rvert}^{\mathcal{K}})$ in the brute force method to $\mathcal{O}({\lvert \mathcal{A} \rvert}\times{\mathcal{K}})$ in the proposed method. Here, the notation $\lvert \mathcal{A} \rvert$ represents the cardinality of the power space of IoT nodes.
    \begin{figure}[h]
    \centerline{\includegraphics[width=0.6\textwidth]{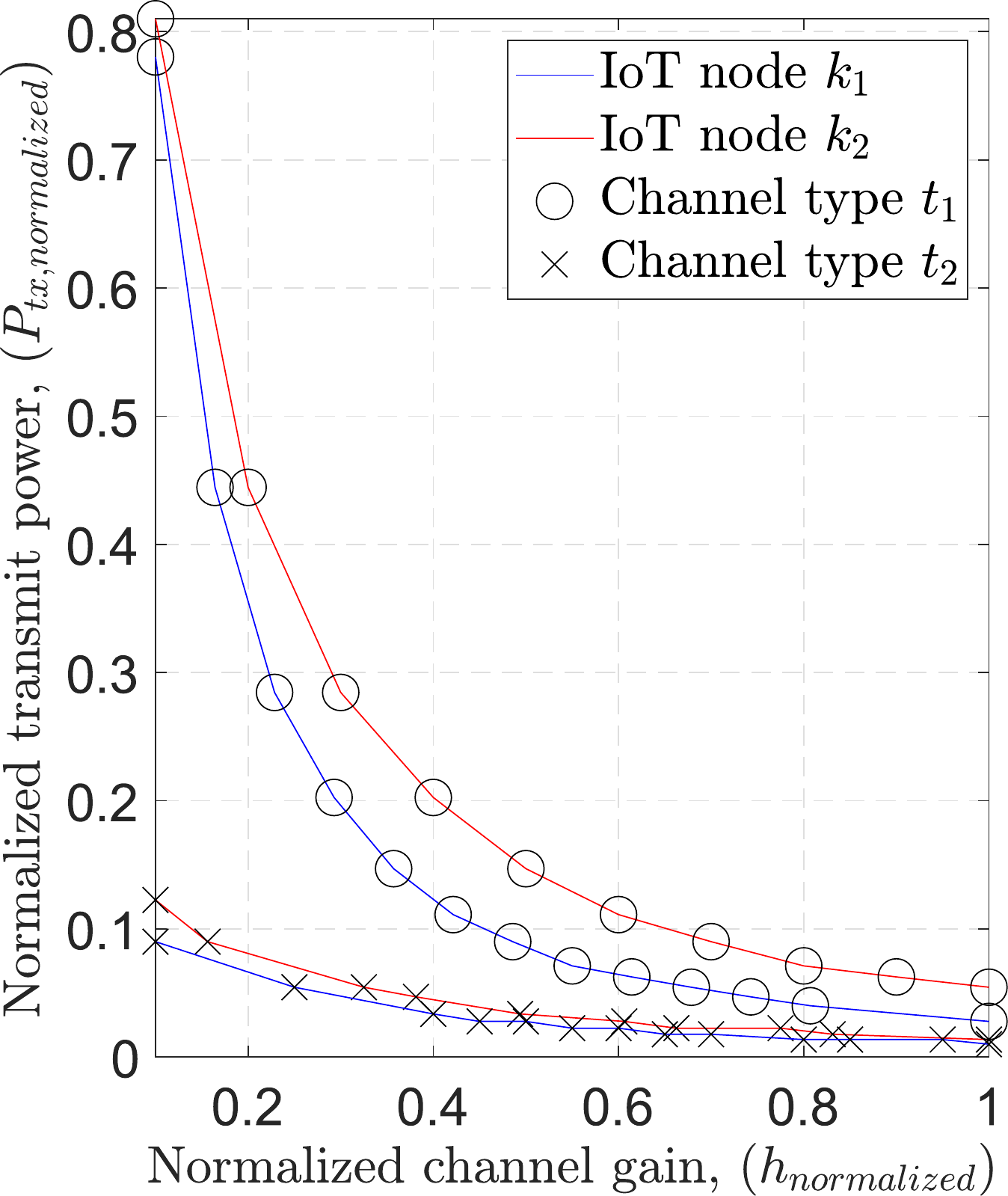}}
    \caption{Variation ($P_{tx, normalized}$ v/s $h_{normalized}$) of optimal transmit Power of IoT nodes $k_{1}$ and $k_{2}$ under channel types $t_{1}$ and $t_{2}$ with normalized channel gain.}
    \label{fig:pow_channel}
    \end{figure}

\figurename~\ref{fig:pow_channel} portrays the variation of optimal transmit power with the channel gain at each time instant for various IoT nodes, considering different channel types. Specifically, the transmit power of each IoT node systematically reduces as the channel gain increases, thereby validating efficiency of the proposed Bayesian approach. In addition, the firing power level of each node increases when transmitting through channels with low-quality gains, and conversely, reduces for channels with high-quality gains. Channel states of type $t_{2}$ exhibit higher levels compared to type $t_{1}$ for both IoT nodes, which consequently receive lower transmit power vectors with a higher gap between each type. This remarkable power difference is influenced not only by the advantages stemming from the channel gain but also by the interplay among different types of IoT nodes. Such interplay can be characterized through the prior probabilities of the channel distribution. To clarify, in this specific case, the game exhibits a stronger inclination towards type $t_{2}$, as it has a higher prior probability of $0.8$, which is intended to capture more of its attributes and converge towards the optimal transmit power. Conversely, a lower prior probability of $0.2$ is allocated to type $t_{1}$, which consequently diminishes its impact on the convergence towards the optimal transmit power. As a consequence, we can utilize probabilistic means to regulate the optimal transmit power strategies for resource allocation among IoT nodes.

    \begin{figure}[t]
    \centerline{\includegraphics[width=0.6\textwidth]{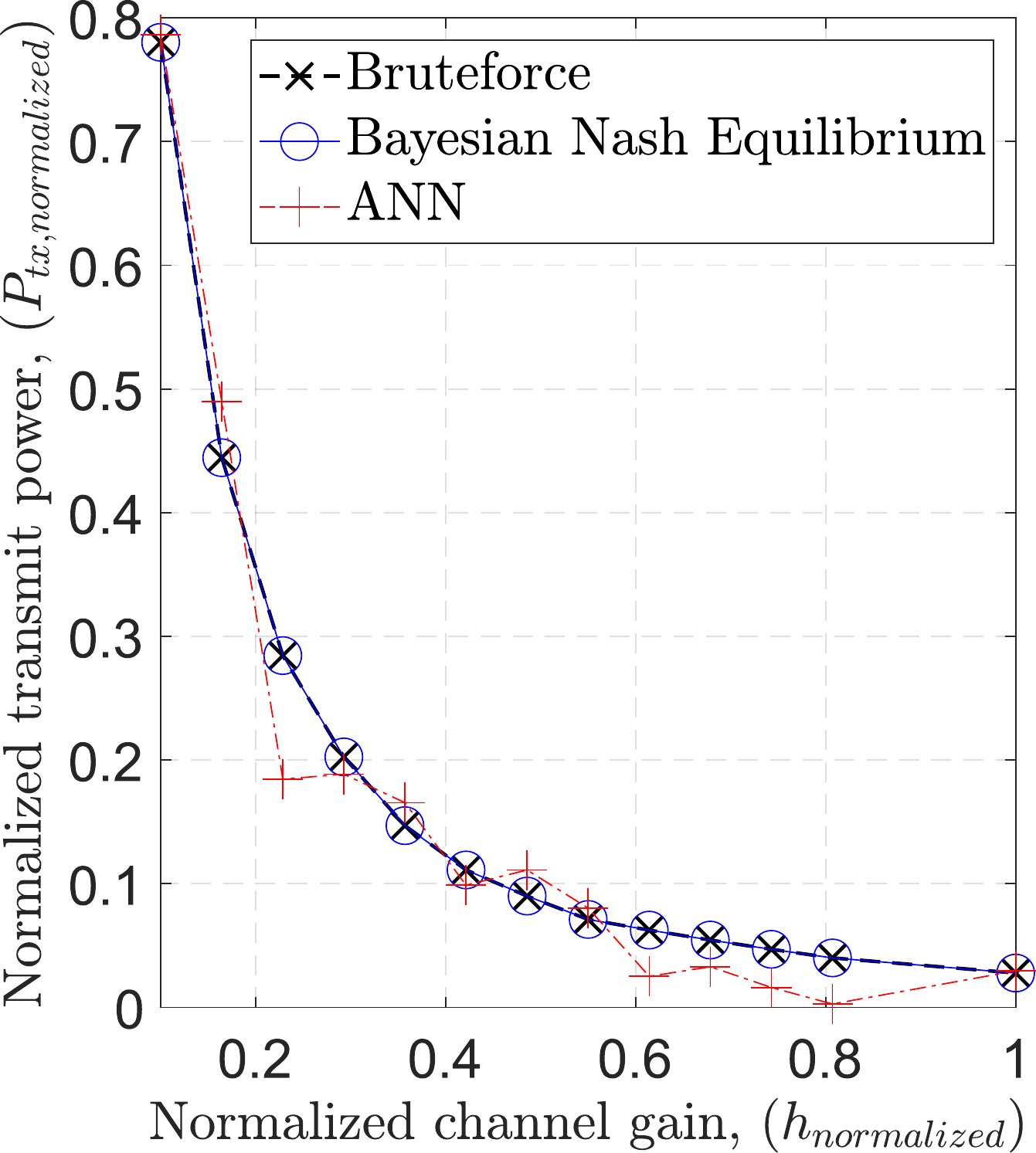}}
    \caption{Variation ($P_{tx, normalized}$ v/s $h_{normalized}$) of optimal transmit Power of IoT nodes $k_{1}$ and $k_{2}$ under channel types $t_{1}$ and $t_{2}$ with normalized channel gain for different techniques.}
    \label{fig:benchmarks}
    \end{figure}
    
\figurename~\ref{fig:benchmarks} depicts variation of node transmit power with channel gain for three different approaches, namely BNE, Bruteforce, and ANN. 
Results confirm that the Bayesian game model aligns well with Bruteforce. The ANN model maintains prediction MSE below 0.007. However, ANN's performance is inferior to the first two methods. The Bayesian approach offers a significant advantage in that it enables the optimization of power consumption under multiple prior probabilistic states. On the other hand, machine learning approaches may encounter significant challenges in dealing with the mixture of prior beliefs due to the increased randomness of the objective function. Although the prediction capability of an ANN model can be enhanced by utilizing a larger training dataset, such improvement is only limited to a specific range of physical parameters. In other words, prediction models with high capacity from the training dataset are only suitable for specific environmental domains and may not perform dynamically. As a result, traditional game theory models can provide a better solution for calculating the optimal power consumption in an adaptive manner under incomplete information of the IoT architecture. 
    \begin{figure}[ht]
    \centerline{\includegraphics[width=0.6\columnwidth]{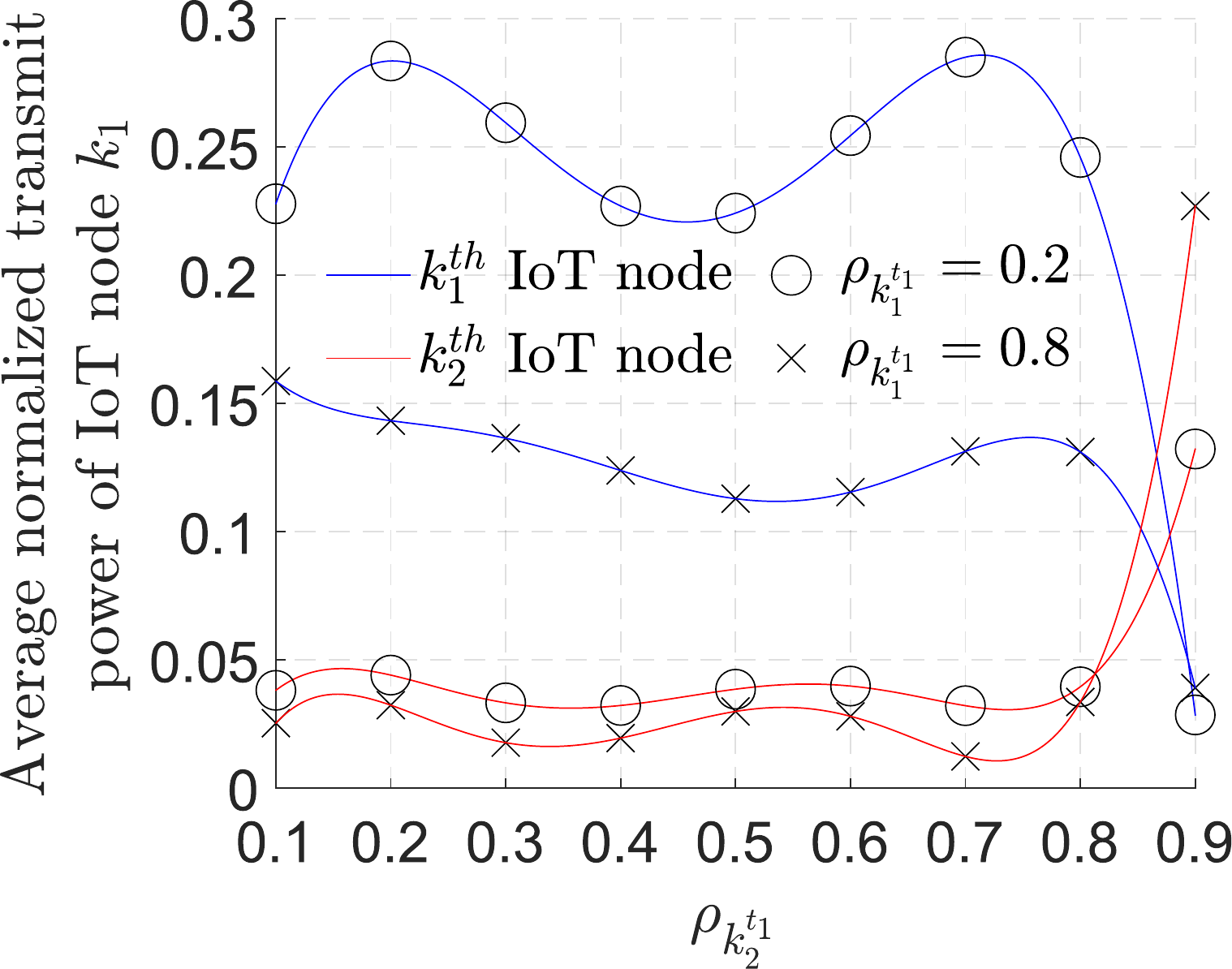}}
    \caption{($\rho_{k_{i}}^{t_{1}}=$ prior probability of $t_{1}^{th}$ channel type of $k_{i}^{th}$ IoT node), Comparison of average optimal normalized transmit power of IoT node $k_{1}$ v/s prior belief of type $t_{1}$ of IoT node $k_{2}$ considering
    prior beliefs of type $t_{1}$ of IoT node $k_{1}$.
    }
    \label{fig:avgpow_prob}
    \end{figure}

\figurename~\ref{fig:avgpow_prob} presents the average of total transmit power for node k under various types and prior beliefs, including the prior belief of node $k_{2}$ under type $t_{1}$. In Bayesian game theory, all nodes possess knowledge about the prior probability distribution solely for the purpose of calculating the utility matrix, and it is not mandatory to calculate individual prior probabilities of each node. Nevertheless, by treating it as a perfect information game, we can gain valuable insights into node interactions within the game mode. Here, lower prior probabilities of type $t_{1}$ for IoT node $k_{2}$ are represented on the x-axis ranging from $0.1$ towards $0.8$,

\begin{itemize}
\item The average transmit power of node $k_{1}$ is lower under type $t_{2}$ than under type $t_{1}$ itself and vice versa.
\item  And, for a specific type of player $k_{1}$, the average transmit power is lower when the prior belief of the respective type is higher, compared to when the prior belief of the type is lower.
\end{itemize}
Next, for the higher prior belief of type $t_{1}$ for node $k_{2}$ which is represented on the x-axis with values greater than  $0.8$, 
\begin{itemize}
\item The average transmit power of node $k_{1}$ is observed to have lower values under type $t_{1}$ than under type $t_{2}$ itself and vice versa.
\item  And, it appears that for a specific type of node $k_{1}$, the average transmit power increases with the higher prior belief of the respective type, than the lower prior belief of the type itself.
\end{itemize}
Therefore, the proposed Bayesian approach exhibits a notable capability to optimize individual transmit powers under varying degrees of knowledge on their neighboring devices' channels, as considered in this work in terms of prior probabilities.

\section{Conclusion}
The primary focus of this paper is to address the transmit power optimization problem of power-constrained IoT devices within massive and dense heterogeneous networks. We propose the utilization of the Bayesian game theory approach to identify the optimal power strategies for each node. In addition, we consider the probability distribution of the channel as common information for all nodes. We formulate simulation-based results to visually demonstrate the existence and uniqueness property of BNE which contributes towards reduction in computational complexity. Simulation results also verify the effectiveness of our proposed approach, which outperforms the traditional ANN approach and is shown to be consistent with the brute-force based results. Finally, we showcase the ability to analyze the interaction among different transmit signals, even when dealing with incomplete information. 

\bibliographystyle{IEEEtran}
\bibliography{IEEEabrv,ref}

\end{document}